\documentstyle[12pt]{article}
\input epsf
\def\be{\begin{equation}}
\def\ee{\end{equation}}
\def\l{\label}
\def\refe#1{(\ref{#1})}

\def\eg{{\it e.g.}}
\def\ie{{\it i.e.}}

\def\LEP2{{LEPII}}
\def\mg{$m_{3/2}$}
\def\th{$\theta$}

\def\npb#1#2#3{    {\it Nucl. Phys. }{\bf B #1} (19#2) #3}

\def\plb#1#2#3{    {\it Phys. Lett. }{\bf B #1} (19#2) #3}

\def\zpc#1#2#3{    {\it Zeit. f\"ur Physik }{\bf C #1} (19#2) #3}

\def\ibid#1#2#3{   {\it ibid. }{\bf #1} (19#2) #3}

\begin{document}

\baselineskip=14pt

\begin{center}

\null
\vskip-1.5truecm
\rightline{IC/96/134}
\vskip1truecm
United Nations Educational Scientific and Cultural Organization\\
and\\
International Atomic Energy Agency\\
\medskip
INTERNATIONAL CENTRE FOR THEORETICAL PHYSICS\\
\vskip1.5truecm
{\bf SUPERSYMMETRIC SUM RULES\\ IN MINIMAL SUPERSTRING
UNIFICATION\footnote{\normalsize  Submitted to Journal of Physics G.}}\\
\vspace{1.2cm}
Shaaban Khalil\footnote{\normalsize Permanent address:
Ain Shams University, Faculty of Science, Department of
Mathematics, Cairo, Egypt.}\\
International Centre For Theoretical Physics, Trieste, Italy.\\
\end{center}
\vspace{0.5cm}
\centerline{ABSTRACT}
\baselineskip=22pt
\medskip

The Minimal superstring unification, assuming orbifold compactification,
provides interesting and rather detailed implications on physics at
low energy. The interesting feature of this model is that the masses of the
spectrum are related since all of them are functions of only two parameters:
the goldstino angle \th\ and the gravitino mass \mg\ . This fact will
help us in studying the modification of the supersymmetric magnetic
moments sum rules which are
very sensitive to the supersymmetry breaking. We write these rules in case
of exact supersymmetry in a form close to the supersymmetric mass relations,
namely $ \sum_J (-1)^{(2J)} (2 J) A_J=0$, where $A_J$ is the anomalous magnetic
moment of the spin $J$ particle. We show that the anomalous
magnetic moments of the W-boson and the gauginos can be written as functions
of \th\ and \mg\ . Then we obtain a  modified version of the
supersymmetric magnetic moment sum rule in the context of the  minimal
superstring unification.
\vspace{1cm}
\begin{center}
MIRAMARE -- TRIESTE\\
July 1996
\end{center}

\vfill\eject

\section{Introduction}

The anomalous magnetic transitions among members of a vector or higher spin
supermultiplet are related by model independent sum rules~\cite{ferrara}. These
rules reduce to
$g_{1/2}=2$ for chiral multiplets and to $ g_{1/2}=2 +2 h $ , $ g_1=2 + h$
for vector
multiplets, where $g_j$ is the gyromagnetic ratio of a given spin $j$
particle as
defined  by $ \mu_j= \frac{e}{M} g_j J $  and h is a real number characterizing
the magnetic transition between the spin-0 and spin-1 states.

The relevant question that has been addressed in many
places, \eg\ ~\cite{culatti} and ~\cite{khalil}, is what is the impact of
the breaking of supersymmetry on these rules.
Clearly, the problem of modifying these sum rules is a problem of SUSY breaking
since
the anomalous magnetic moment depends on the supersymmetric spectrum which
is
determined in terms of the SUSY soft breaking terms. In the Minimal
Supersymmetric
Standard Model (MSSM) with supersymmetry broken explicitly but softly by a
universal
mass $\tilde{m}$ for all scalar particles, the total contribution of the
anomalous
magnetic moments of the W-boson $\Delta K_{WW} $, the anomalous magnetic
moments of
the charginos $a_{\omega_1}$ and $a_{\omega_2}$ and the magnetic transition
between
the spin-1 and spin-0 states in a vector multiplet  $\Delta K_{WH} $ has been
considered in ~\cite{culatti}. However, in this case the sum rules
\be
 \Delta K_{WW} = a_{\omega_1}= a_{\omega_2}= \Delta K_{WH}
\l{rules}
\ee
results to be badly broken without any interesting functional relation among the
four quantities. In Ref.~\cite{khalil}, we pursued the same strategy in
the case of
spontaneous breaking of the global N=1 supersymmetry. The hope was that
the spontaneous nature of supersymmetry breaking could guarantee the
survival of some interesting
relations among the various transition magnetic moments. We considered a SUSY
spontaneous breaking realized a`la Fayet-Iliopoulos in the realization of
Ref.~\cite{barbieri}. In that model the following mass splitting is obtained:
\begin{eqnarray}
\Delta m_{gauge}^2 = \mu^2 , \\
\Delta m_{matter}^2 = {1\over 4} \mu^2
\l{spliting}
\end{eqnarray}
where $\Delta m_{gauge}$ and $\Delta m_{matter}$ denote the mass differences
between the fermionic and bosonic components in the vector and scalar multiplet
respectively, and $\mu$ is proportional to the v.e.v of the
Fayet-Iliopoulos term. Our result showed that the sum rules \refe{rules}
are again badly broken without any surviving clear pattern.

	The investigation of this problem in the full realistic case of
spontaneous  broken N=1 supergravity is the aim of this paper. The most
popular way to break SUSY is to assume that the flatness of moduli and
dilaton directions of the effective potential are lifted by non-perturbative
dynamics and that SUSY breaking arises from the non-vanishing VEV's of the
F-term of modulus $T$ and/or dilaton $S$ supermultiplets. The soft terms
become, in general, functions of the the gravitino mass $m_{3/2}$ and the
goldstino angle $\theta$~\cite{ibanez}. We have shown that the scheme
of minimal superstring unification provides interesting and rather
detailed implications on physics to be tested in \LEP2\ . In this paper we
study the modification of the  supersymmetric sum rules in the minimal
superstring unification.

The outline of the paper is as follows. In Section 2 we
present a brief review of the minimal superstring unification, and we determine
the masses of the supersymmetric
spectrum which will be needed for calculating the anomalous magnetic moment.
In Section 3, we calculate the anomalous magnetic moment of the spin-1 gauge
boson and spin-1/2 partner, respectively showing that they are given in
terms of
only
two parameters: the goldstino angle \th\ and the gravitino mass \mg\ .
Then we discuss
the modification of the supersymmetric sum rules in the context of the minimal
superstring unification. In Section 4, we present our conclusions.
\section{{\large{\bf The Minimal Superstring Unification Model}}}
In this section, we give a brief review of the construction of the soft SUSY
breaking terms in the minimal superstring unification models ~\cite{ross}, where
the orbifold compactification with large threshold correction is assumed.
The soft breaking terms have the form:\\
The scalar masses are
\begin{equation}
m^2_i = m^2_{3/2}(1 + n_i \cos^2\theta)
\l{scalar}
\end{equation}
where $n_i$ are the modular weights which are given by
$$n_{Q_L}=n_{D_R}=-1,\hspace{0.5cm} n_{u_R}=-2,\hspace{0.5cm}
n_{L_L}=n_{E_R}=-3,\hspace{0.5cm} n_{H_1}=-2,\hspace{0.5cm} n_{H_2}=-3$$
as was explained in ~\cite{masiero}.
The gaugino masses are
\begin{eqnarray}
 M_1 = \sqrt{3} m_{3/2} \left( \sin\theta - 0.02 \cos\theta \right)\\
 M_2 = \sqrt{3} m_{3/2} \left( \sin\theta + 0.06 \cos\theta \right)\\
 M_3 = \sqrt{3} m_{3/2} \left( \sin\theta + 0.12 \cos\theta \right)
\end{eqnarray}
The trilinear coupling is
\begin{equation}
A_t = - m_{3/2}(\sqrt{3}  \sin\theta - 3 \cos\theta)
\end{equation}
The $A_t$ term is the only term relevant to the radiative symmetry breaking
since we assume that the only top-Yukawa coupling is nonvanishing:
Finally, the bilinear coupling is
\begin{equation}
B = m_{3/2}(-1 - \sqrt{3}  \sin\theta + 2 \cos\theta)
\l{b}
\end{equation}
Given the boundary conditions in equations \refe{scalar} to \refe{b} at the
compactification scale $M_S=3.6 \times 10^{17}$ GeV, we have to
determine the
evolution of the couplings according to their renormalization group
equation (RGE) to finally compute the mass spectrum of the SUSY particles
at the weak scale.
For a detailed discussion of these points see Ref.~\cite{thesis}.\\
At the weak scale we find
\be
A_t  =  m_{3/2} \left[ ( 3.817 \cos \theta + 5.739 \sin \theta)- r
( 3.446 \cos \theta + 2.495 \sin \theta)\right],
\l{at}
\ee
\be
B = m_{3/2} \left[ ( -1 +2.057 \cos \theta -0.64137 \sin \theta) - r
( 1.723 \cos \theta + 1.24767 \sin \theta) \right],
\ee
\be
m_{H_1}^2= m_{3/2}^2 \left ( 1-1.85 \cos^2 \theta +
0.182 \cos\theta \sin\theta + 1.67 \sin^2\theta \right),
\ee
and
\begin{eqnarray}
m_{H_2}^2 & = & m_{3/2}^2 \left(1-3.14 \cos^2\theta+0.18 \cos\theta \sin\theta+
1.67\sin^2\theta \right)
	\nonumber\\
	& + &  m_{3/2}^2 r \left( -1.5 -3.11 \cos^2\theta-11.58 \cos\theta \sin\theta -
16.44 \sin^2\theta \right)
	\nonumber\\
	& + & m_{3/2}^2 r^2 \left(5.94 \cos^2\theta +8.6 \cos\theta \sin\theta +
3.22 \sin^2\theta) \right).
\end{eqnarray}
where $$ r= \frac{Y_t(0)}{Y_t(0) + 5 \times 10^{-4}}$$
 and $Y_t= \frac{\lambda_{t}^2}{(4\pi)^2}$ and $Y_t(0)$ is the t-Yukawa
coupling
at $M_S$.\\
The  electroweak symmetry breaking requires the following conditions among the
renormalized $\mu_1^2$, $\mu_2^2$ and $\mu_3^2$ quantities: \\
\be
\mu_1^2 +\mu_2^2 > 2\mu_3^2, \hspace{1cm} \vert \mu_3 \vert ^4 > \mu1_1^2
\mu_2^2.
\ee
and
\be
\mu^2 = \frac{ m_{H_1}^2 -m_{H_2}^2 \tan^2\beta}{\tan^2\beta - 1}
- \frac{M_Z^2}{2},
\hspace{1.5cm} \sin 2\beta= \frac{-2  B \mu }{m_{H_1}^2+m_{H_2}^2+ 2\mu^2 }
\l{minimization}
\ee
where $\tan\beta= {\langle H_2^0 \rangle}/{\langle H_1^0 \rangle}$. Using
equations
\refe{at}-\refe{minimization} we find that $\mu$ and $\tan \beta $ are
given in terms
of the goldstino angle and the gravitino mass. Fig.1 shows $\tan \beta$
as a function of \th\ for different values of \mg\ .
Thus all the low energy
quantities can be determined  in
terms of only the gravitino mass \th\ and the gravitino angle \mg\ .
As we have explained in ~\cite{masiero} a further constraint on the
parameter space is entailed by the demand of
colour and electric charge conservations. In particular, the latter
conservation yields the most powerful constraint \cite{savoy} and we find
that  \th\ is  limited approximately by
$\theta \in [0.98,2]$ rad , and \mg\ is larger than 55 GeV.

	As we will see in the next section the mass of the squarks,
sleptons and charginos are needed to calculate the anomalous magnetic
moment of the W-boson and  the gauginos $\omega_1$ and $\omega_2$. Here we
will show explicitly the dependence
of these masses on the gravitino mass \mg\ and the goldstino angle \th\ .
It is well
known that the complete expressions for the first two generations squark  and
slepton
mass parameters are given by
\be
m^2_{\tilde{f}}= m^2 + \sum_{j=1}^{3} f_j M_j^2 +Y_{f} \tilde{\alpha_1} S(t)
+ ( T^3_f -Q_f \sin^2 \theta_W) M^2_Z \cos^2 \beta
\l{msu}
\ee
where the sum is over the gauge groups $U(1)$, $SU(2)$ and $SU(3)$ and
$$ f_j= \frac{c_j(f)}{b_j} [1-\frac{1}{(1-\frac{\alpha_{String}}{2\pi}
b_j t)^2}],$$
with $b_j = ( 11,1,-3)$ for $U(1)$, $SU(2)$ and $SU(3)$ respectively and
$c_j$ is $\frac{N^2 -1 }{N}$(0) for
the fundamental (singlet) representation of SU(N) and $Y^2$ for $U(1)_Y$.
>From equation \refe{msu} we  find that the masses of the superpartners are
functions of \th\ and \mg\ . For example the mass of the superpartner of
the up quark, $m_{\tilde{u}}^2$, is given by
\begin{eqnarray}
m_{\tilde{u}}^2 & = & m_{3/2}^2\left(1-0.7 \cos^2\theta +5.53\cos\theta
\sin\theta + 23.829 \sin^2\theta \right)
	\nonumber\\
	& - & m_{3/2}^2 r \left(0.5+1.037 \cos^2\theta +3.86 \cos\theta
\sin\theta +5.48 \sin^2\theta \right)
	\nonumber\\
	&+& m_{3/2}^2  r^2\left(1.979 \cos^2\theta +2.866
\cos\theta\sin\theta +1.037 \sin^2\theta \right)
	\nonumber\\
	&+& M_Z^2 \cos 2\beta (1/2 -2/3 \sin^2 \theta_W).
\end{eqnarray}
Similar formulae for the mass of the superpartner of the down
quark $m_{\tilde{d}}^2$ , the mass of selectron $m_{\tilde{e}}^2$
and the mass of sneutrino $m_{\tilde{\nu}}^2$ can be obtained.

The squark mass spectrum of the third generation is more complicated
for two reasons:
(1) the effects of the third generation Yukawa couplings need not be
negligible.\\
(2) there can be substantial mixing between the left and the right top squark
fields so that they are not mass
eigenstates.\\
Let us keep the top Yukawa coupling only and neglect the others.
Therefore, the relations for the mass of $\tilde{b}$, $\tilde{\tau}$ will
be as for the first two generations. The masses of the stop are given by:
\be
m_{1,2}^2= \frac{1}{2} ( m_{LL}^2+m_{RR}^2 \pm ((m_{LL}^2-m_{RR}^2)^2+
4 m_{RL}^4)^{\frac{1}{2}})
\l{stop}
\ee
where $m_{LL}$, $m_{RR}$ and $m_{RL}$ are as defined in Ref.~\cite{Lopez}.
>From equation \refe{stop} we can easily see that the mass of the stop
quarks are
also
functions of only \mg\ and \th\ . Fig. 2 shows the relation between light
stop and the \mg\ and \th.\\
	Finally, we are also interested in the mass of the charginos
$\omega_1$ and  $\omega_2$. They are given by
\begin{eqnarray}
M_{2,1}^2 & = & 1/2 ( M_2^2 +\mu^2 +2M_W^2
	\nonumber\\
	& \pm & \sqrt{(M_2^2-\mu^2)^2 + 4
M_W^4\cos^2 2\beta + 4M_W^2 (M_2^2 +\mu^2 + 2M_2\mu \sin 2\beta)}).
\end{eqnarray}
It is also clear that $M_{2,1}^2$ are functions of \th\ and \mg\ . Fig.3
shows the mass of the lightest chargino as a function of the goldstino angle
\th\ for different values of the gravitino mass \mg\ .
\section{{\large{\bf Supersymmetric Sum Rules In the Minimal\\
 Superstring Unification}}}
	In supersymmetric theories it was shown that the existing sum rule holds
at any order in perturbation theory implying that the anomalous magnetic
moment of
spin 1/2 particle in chiral supermultiplet is identically zero ( i.e.
$g_{1/2}=2$),
and
 \be
 \Delta K_{WW} = a_{\omega_1}= a_{\omega_2}= \Delta K_{WH}
\ee
for the vector multiplet. These rules have been verified in case of massles
ordinary fermions ~\cite{van} and
for massive fermions~\cite{masiero1} and ~\cite{culatti}, where we have
$m_b=0$ with fixed ratio $(\frac{m_W}{m_t})^2 =\alpha, $
\begin{equation}
\Delta K_{WW}^{q\tilde{q} l\tilde{l}} = a_{\omega_1}^{q\tilde{q} l\tilde{l}} =
a_{\omega_2}^{q\tilde{q} l\tilde{l}} = \Delta K_{WH}^{q\tilde{q} l\tilde{l}} =
\frac{-g^2}{32\pi^2} G(\alpha )
\end{equation}
with
\begin{equation}
G(\alpha )=\frac{2}{\alpha^2} [3\alpha +(3-2\alpha )\ln (1-\alpha )].
\end{equation}
This function has $limit_{\alpha\rightarrow 0} G(\alpha) =1$ and
so reproduces the case of negligible $m_W$.  Using the value of the top
quark  mass $m_t=174$, we find that  $\alpha=0.211389 $  and
\be
\Delta K_{WW}^{q\tilde{q} l\tilde{l}} = a_{\omega_1}^{q\tilde{q}
l\tilde{l}} =
a_{\omega_2}^{q\tilde{q} l\tilde{l}} = \Delta K_{WH}^{q\tilde{q}
l\tilde{l}} =
\frac{-g^2}{16\pi^2} \times 0.495082.
\l{exact}
\ee
	The work of Refs.~\cite{culatti} and ~\cite{khalil} showed that
these sum rules are very sensitive to supersymmetry breaking, and even
in a very simplified version of a theory with supersymmetry is broken by
a universal mass $\tilde{m}$ for all the scalar, the sum rules \refe{rules}
is badly broken.

The breaking of supersymmetry in the minimal superstring unification is a
consistent way and we have determined explicitly the
allowed region in the parameter space where all the experimental and
theoretical constraints are satisfied  ~\cite{masiero}. The interesting feature of this
model is that the masses of the supersymmetric particles, as we
explained in the previous section, are related since all of them are
functions of \th\ and
\mg\ . This arises again the hope of modifying these sum rules.

	We know that, in supersymmetric theories, strong mass relations hold
\be
STr M^2= \sum_J (-1)^{(2J)} (2 J +1) M_J=0	
\l{relation1}
\ee
and in the case of spontaneously broken local supersymmetry these relations get
modified
to
\be
STr M^2= \sum_J (-1)^{(2J)} (2 J +1) M_J= 2(N-1)m_{3/2} -2 R_j^i F_i F^j
\l{relation2}
\ee
with
\be
R_j^i= [\log det ( G_k^l)]_j^i\ , \hspace{1.5cm} F_i= exp(-G/2) (G^{-1})^j_i G_j
\ee
where $G$ is the K\"ahler potential, and $G^i$ is the derivative of $G$. In
Ref.
~\cite{ibanez}
the F-term of the dilaton field $S$ and modulus field $T$, which are the
only fields
contributing in breaking supersymmetry, are parameterized as follow
\be
(G_{0S}^S)^{1/2} F_0^S = \sqrt{3} m_{3/2}  sin \theta \ee
\be
(G_{0T}^T)^{1/2} F_0^T = \sqrt{3} m_{3/2}  cos \theta
\ee
We have shown that in this scheme of supersymmetry breaking
all the supersymmetric masses are functions of \th\ and \mg\ . Then in the
minimal
superstring unification the relation \refe{relation2} takes the form
\be
STr M^2= \sum_J (-1)^{(2J)} (2 J +1) M_J= f(m_{3/2}, \theta )
\l{relation3}
\ee
and in case of pure dilaton \ie\ $\theta = \pi/2$ it becomes
\be
STr M^2= \sum_J (-1)^{(2J)} (2 J +1) M_J= 2 (N-1) m_{3/2}.
\l{relation4}
\ee
We find that the supersymmetric magnetic moment sum rules can be written in a
form close to
the relations in Eq. \refe{relation1}, namely
\be
\sum_J (-1)^{(2J)} (2 J) A_J=0
\ee
where $A_J$ is the anomalous magnetic moment of the spin $J$ particle. This
simple relation gives the supersymmetric sum rules given in Ref.
\cite{ferrara} in both cases of chiral and vector supermultiplet. We can
see this easily: in case of chiral supermultiplet
this relation leads to $ A_{1/2}=0$ and in the case of vector multiplet it
leads to
$$  2 \Delta K_{WW} - a_{\omega_1} - a_{\omega_2} = 0 $$
and since in exact supersymmetry $a_{\omega_1}=a_{\omega_2}$ we find
$$ \Delta K_{WW} = a_{\omega_1} = a_{\omega_2}  $$
which is given in Eq. \refe{rules}.

	The question now is how will this relation be modified in case
supersymmetry
is broken in the context of the minimal superstring unification and in
particular in case of
the pure dilaton? Can we get a relation similar to the modified mass
relation in  Eq. \refe{relation4}? To answer these questions, we will
calculate the total contribution of
the anomalous magnetic moment of the W-boson $ \Delta K_{WW}$ , and the
anomalous  magnetic moment of the charginos $a_{\omega_1}$ and
$a_{\omega_2}$. We will concentrate on  the case of vector
supermultiplet, since in the case of chiral supermultiplet, as we explained in
~\cite{khalil}, the anomalous magnetic moment of the fermionic member
is different from zero and depends only on the mass of the superpartner, \ie\
in the minimal superstring unification  it depends only on the \th\ and
\mg\ . The result of the one loop  contributions to the anomalous magnetic
moment of the W-boson, $\Delta K_{WW}$ is given by:\\
For first two quark generations, $$ \Delta K_{WW}= -\frac{g^2}{16\pi^2}.$$
while for the third quark generation, $$ \Delta K_{WW}= -0.95347 \times
\frac{g^2}{16\pi^2}.$$
Then the quark contribution to the anomalous magnetic moment of W-boson is
given by
\be
\Delta K_{WW}(q) = -1.4767  \times \frac{g^2}{16\pi^2}
\ee
The squark contribution to the anomalous magnetic moment of W-boson is
given by
\begin{eqnarray}
\Delta K_{WW} & = & \frac{g^2}{16\pi^2} \int_0^1 dx\,
\frac{(x^3-x^2)(\tilde{b}-\tilde{a} -1 +2x)}{\tilde{b} x+\tilde{a}(1-x)-
x(1-x)}
	\nonumber\\
    & + & \frac{2g^2 }{16\pi^2} \int_0^1 dx\,
\frac{(x^3-x^2)(\delta{a}-\tilde{b} -1 +2x)}{\tilde{a}x+\tilde{b}(1-x)-x(1-x)}
\end{eqnarray}
for one generation. The first two squark generations are identically the
same, since  $\tilde{a}=(\frac{m_{\tilde{u}}}{m_W})^2$ and
$\tilde{b}=(\frac{m_{\tilde{b}}}{m_W})^2$,
while the third squark generation contribution is different because $\tilde{a}$
is given by  $\tilde{a}=(\frac{m_{\tilde{t}}}{m_W})^2$. As we explained in
the previous section $ m_{\tilde{u}}^2$ and $ m_{\tilde{d}}^2$ are
functions of \mg\ and \th\ and therefore $\Delta K_{WW}$ is also a
function of \mg\ and \th\ . \\
The lepton contribution to $\Delta K_{WW}$ is given by
\be
\Delta K_{WW}(l) = -0.5 \times \frac{g^2}{16\pi^2}.
\ee
The slepton contribution to $\Delta K_{WW}$ is given by
\be
\Delta K_{WW}(\tilde{l}) = 3 \times \frac{g^2}{16\pi^2}\int_0^1 dx\,
\frac{(x^3-x^2)(\tilde{b}-\tilde{a} -1 +2x)}{\tilde{b}
x+\tilde{a}(1-x)-x(1-x)},
\ee
where $\tilde{a}=(\frac{m_{\tilde{\nu_e}}}{m_W})^2$ and
$\tilde{b}=(\frac{m_{\tilde{e}}}{m_W})^2$. It is clear that
$\Delta K_{WW}(\tilde{l})$  is also a function of \th\ and \mg.
In Fig.4  we plot the total contribution of quarks, squarks, leptons and
sleptons to the anomalous magnetic moment of the W-boson,
$\Delta K_{WW}(q, \tilde{q},\l,\tilde{l})$, as a function of the
gravitino mass \mg\ and for different values of the goldstino angle \th\ .\\

The anomalous magnetic moment of the gauginos $a_{\omega_1}$
are given by:
\begin{eqnarray}
a_{\omega_1}&=& \frac{g^2}{16\pi^2} ( 4 \int_0^1 dx\,
\frac{x^2(x-1)}{\tilde{a}x -x(1-x)} -
2 \int_0^1 dx\, \frac{x^2(x-1)}{\tilde{a}(1-x)-x(1-x)}
	\nonumber\\
	& - &2 \int_0^1 dx\, \frac{\tilde{b} x^2(1-x)}{\tilde{b}x-x(1-x)} +
4 \int_0^1 dx\, \frac{\tilde{b} x^2(1-x)}{\tilde{b}(1-x)-x(1-x)} )
\end{eqnarray}
for the first two generations of the quark and squark, and
$\tilde{a}=(\frac{m_{\tilde{u}}}{m_{\omega_1}})^2$ ,
$\tilde{b}=(\frac{m_{\tilde{d}}}{m_{\omega_1}})^2$. While for the third
generation it is given by
\begin{eqnarray}
a_{\omega_1}&= &\frac{g^2}{16\pi^2} ( 2 \int_0^1 dx\,
\frac{x^2(x-1)}{\tilde{a}x -x(1-x)} -
\int_0^1 dx\, \frac{x^2(x-1)}{\tilde{a}(1-x)-x(1-x)}
	\nonumber \\
	& - & \int_0^1 dx\, \frac{\tilde{b}
x^2(1-x)}{\tilde{b}x+a (1-x)-x(1-x)} +
2 \int_0^1 dx\, \frac{\tilde{b} x^2(1-x)}{a x+ \tilde{b}(1-x)-x(1-x)} )
\end{eqnarray}
where $\tilde{a}=(\frac{m_{\tilde{t}}}{m_{\omega_1}})^2$ ,
$\tilde{b}=(\frac{m_{\tilde{b}}}{m_{\omega_1}})^2$
and $a=(\frac{m_t}{m_{\omega_1}})^2$. These quantities are given in
terms of \th\ and \mg\ , and therefore $a_{\omega_i}$ is
a function of \th\ and \mg.
Finally, the lepton contribution is given by
\be
a_{\omega_1}= \frac{g^2}{16\pi^2} (-3 \int_0^1 dx\,
\frac{x^2(x-1)}{\tilde{a}(1-x) -x(1-x)} -
\int_0^1 dx\, \frac{x^2\tilde{b}(1-x)}{\tilde{b}x -x(1-x)}
\ee
Similar formulae for $a_{\omega_2}$ can be obtained by replacing
${m_{\tilde{u}}}$ by ${m_{\tilde{d}}}$ and vice versa, also by changing the
value of the electric charge of u-quark to the negative value of the
electric charge of the d-quark and vice versa.\\

	We are interested in finding the modification of the supersymmetric
sum rules \refe{rules}
due to the breaking of supersymmetry. It is clear that it is difficult to
deduce a relation among the
$\Delta K_{WW}$ and $a_{\omega_i}$ from the above results.
So we will try to use the above results to find the
functions that  depend on the two parameters which $\Delta K_{WW}$ and
$a_{\omega_i}$ depend on, namely,  \th\ and \mg\
and fit all the above results of $\Delta K_{WW}$ and $a_{\omega_i}$  .\\

	Using the Statistics Nonlinear Fit package in Mathematica we have
obtained the functions that  fit the results of $\Delta
K_{WW}$, $a_{\omega_1}$ and $a_{\omega_2}$ as follows:
\be
\Delta K_{WW} = \frac{g^2}{16 \pi^2} \left( -0.495082 -0.02
(\frac{m_{3/2}}{m_W})^2 \cos\theta - 0.2 (\frac{ m_{3/2}}{m_W})^2 \sin\theta
\right) \l{mod1}
\ee
\be
a_{\omega_1} = \frac{g^2}{16 \pi^2} \left(-0.495082 +0.12
(\frac{m_{3/2}}{m_W})^2 \cos\theta + 0.28 (\frac{m_{3/2}}{m_W})^2 \sin\theta
\right) \ee
\be
a_{\omega_2} =\frac{g^2}{16 \pi^2} \left( -0.495082 -0.08
(\frac{m_{3/2}}{m_W})^2 \cos\theta + 0.03 (\frac{m_{3/2}}{m_W})^2 \sin\theta
\right). \l{mod2}
\ee
Equations \refe{mod1}-\refe{mod2} show explicitly that  $\Delta K_{WW}$,
$a_{\omega_1}$ and $a_{\omega_2}$ are functions of \th\ and \mg. Also in the
limiting case of exact supersymmetry ( i.e. \mg\ =0 ) we obtain the
values of $\Delta K_{WW}$, $a_{\omega_1}$ and $a_{\omega_2}$ in equation
\refe{exact}. So that in case of breaking
supersymmetry in the context of minimal superstring unification we have
\be
\sum_J (-1)^{(2J)} (2 J) A_J= \frac{g^2}{16 \pi^2} [-0.08
(\frac{m_{3/2}}{m_W})^2 \cos \theta -0.7 (\frac{m_{3/2}}{m_W})^2 \sin \theta
].  \l{mrule}
\ee
This modified rule gives a relation between the anomalous magnetic
moment among members of a vector supermultiplet in case of breaking
supersymmetry in the context of the minimal superstring unification. It is
also clear that in the case of pure dilaton, where $\theta = \pi/2$, these
modified rules reduce to
\be
\sum_J (-1)^{(2J)} (2 J) A_J= \frac{g^2}{16 \pi^2} -0.07
(\frac{m_{3/2}}{m_W})^2
\ee
which is very similar to the modification of the mass relation in the case of
pure dilaton
\refe{relation4}. Moreover we can get a relation between the anomalous magnetic
moment of the W-boson and the gauginos $\omega_1$ and $\omega_2$
independent of the supersymmetry breaking parameters. In pure dilaton
scenario this relation is as follows:
\be
\Delta K_{WW} + 0.64 a_{\omega_1} + 0.7 a_{\omega_2} = -1.16 \times
\frac{g^2}{16 \pi^2}
\ee
This relation between the anomalous magnetic moment of different particles
within
the same
supermultiplet could be a possible way of testing SUSY theories.
\section{Conclusion}
	We studied the modification of supersymmetric sum rules in the minimal
superstring unification. We determined the mass spectrum of squarks, sleptons
and the charginos to calculate the anomalous magnetic moment of the W-boson
and the charginos. The interesting feature of the minimal superstring
unification  is that all the spectrum is determined in terms of
two parameters.\\
	We obtained supersymmetric magnetic moment sum rules that relate the
 anomalous magnetic moment of the members of a vector supermultiplet, and
 they reduce to the SUSY sum rules defined in ~\cite{ferrara} in the case of
 exact  supersymmetry.\\

\noindent{\Large\bf Acknowledgments}
\vskip0.5truecm
The author would like to thank Professor A Masiero for his guide and
A. Smirnov and F. Vissani for useful discussions. He would also like to
thank Professor S. Randjbar-Daemi, UNESCO and the International Atomic
Energy Agency
 for  hospitality at the International Center for Theoretical
Physics, Trieste.\\

\newpage

\noindent{\Large\bf Figure Captions}
\vskip0.5truecm
\noindent{\bf Fig.\ 1} The values of $\tan \beta$ as a function of the
goldstino angle.\\
{\bf Fig.\ 2} The lightest chargino mass in the region of
interest for \LEP2\ searches, as a function of the goldstino angle.
The horizontal lines correspond to the visibility at \LEP2. While the
vertical line corresponds to the pure dilaton breaking.\\
{\bf Fig.\ 3} The lightest stop mass as a function of the goldstino
angle versus.\\
{\bf Fig.\ 4} The total contribution of quarks, squarks, leptons, and
sleptons to the anomalous magnetic moment of W-boson as a function of the
goldstino angle.\\

\newpage

\begin{figure}
\epsfxsize=\hsize
\epsffile{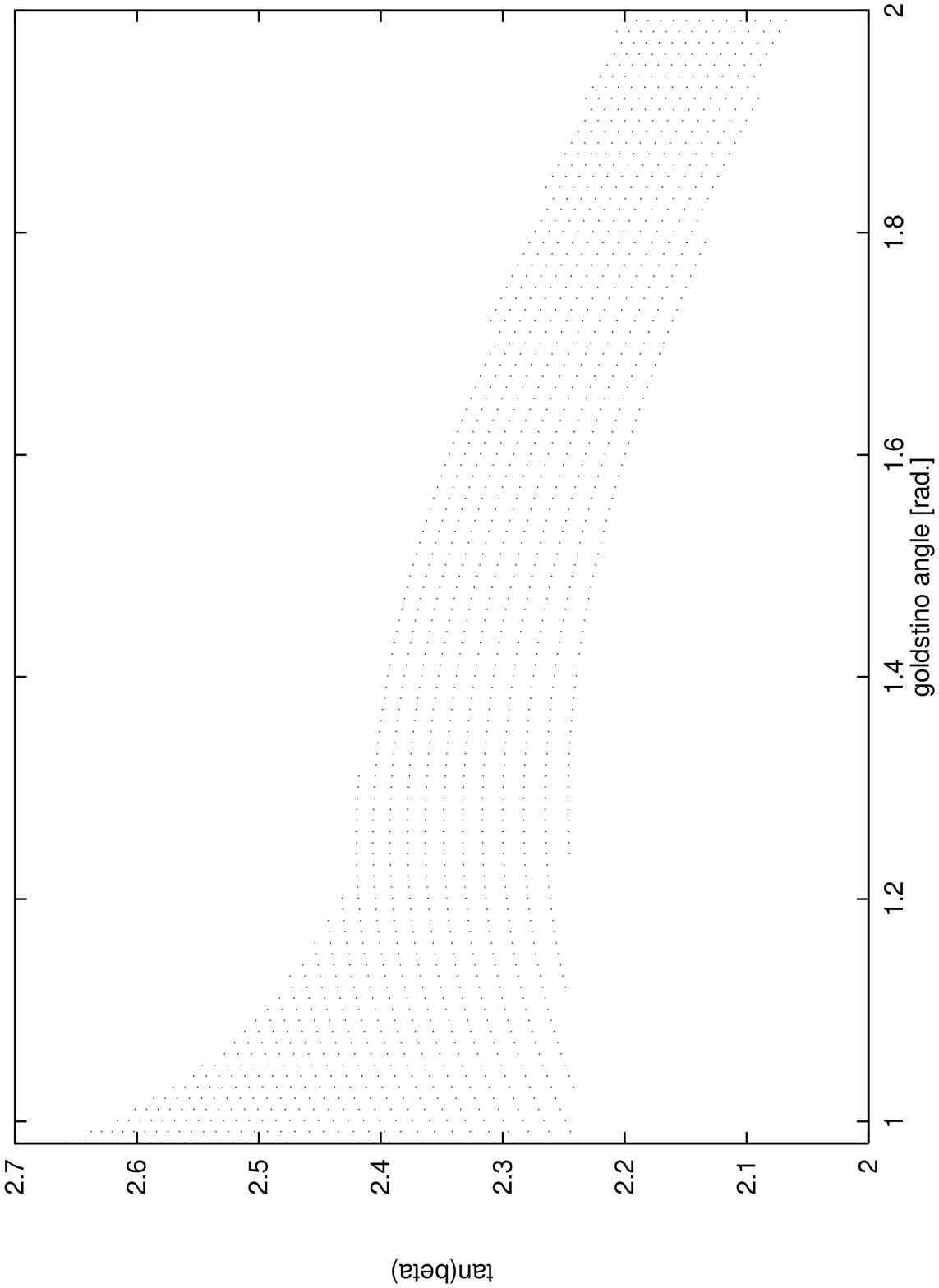}
\end{figure}
\begin{figure}
\epsfxsize=\hsize
\epsffile{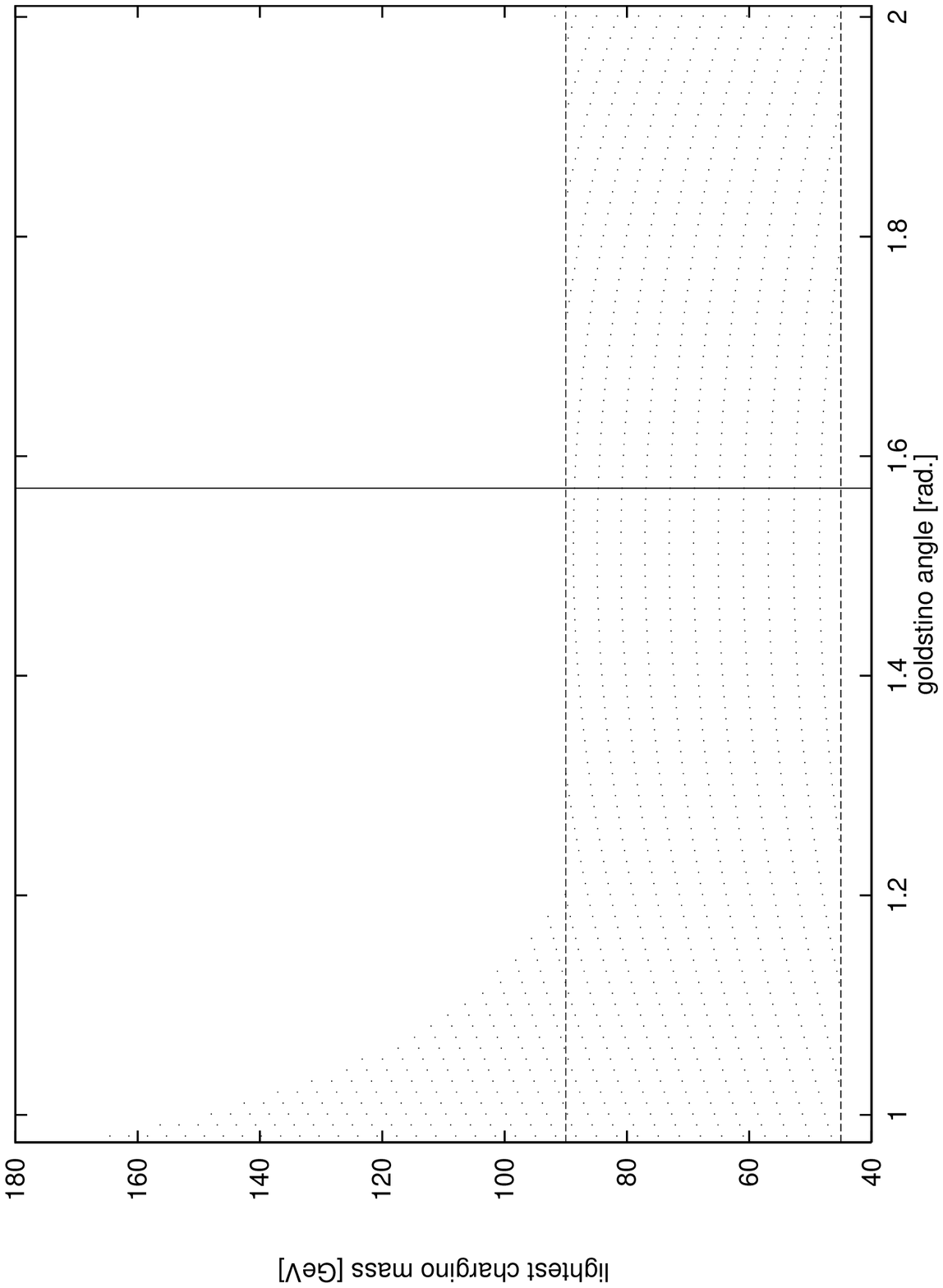}
\end{figure}
\begin{figure}
\epsfxsize=\hsize
\epsffile{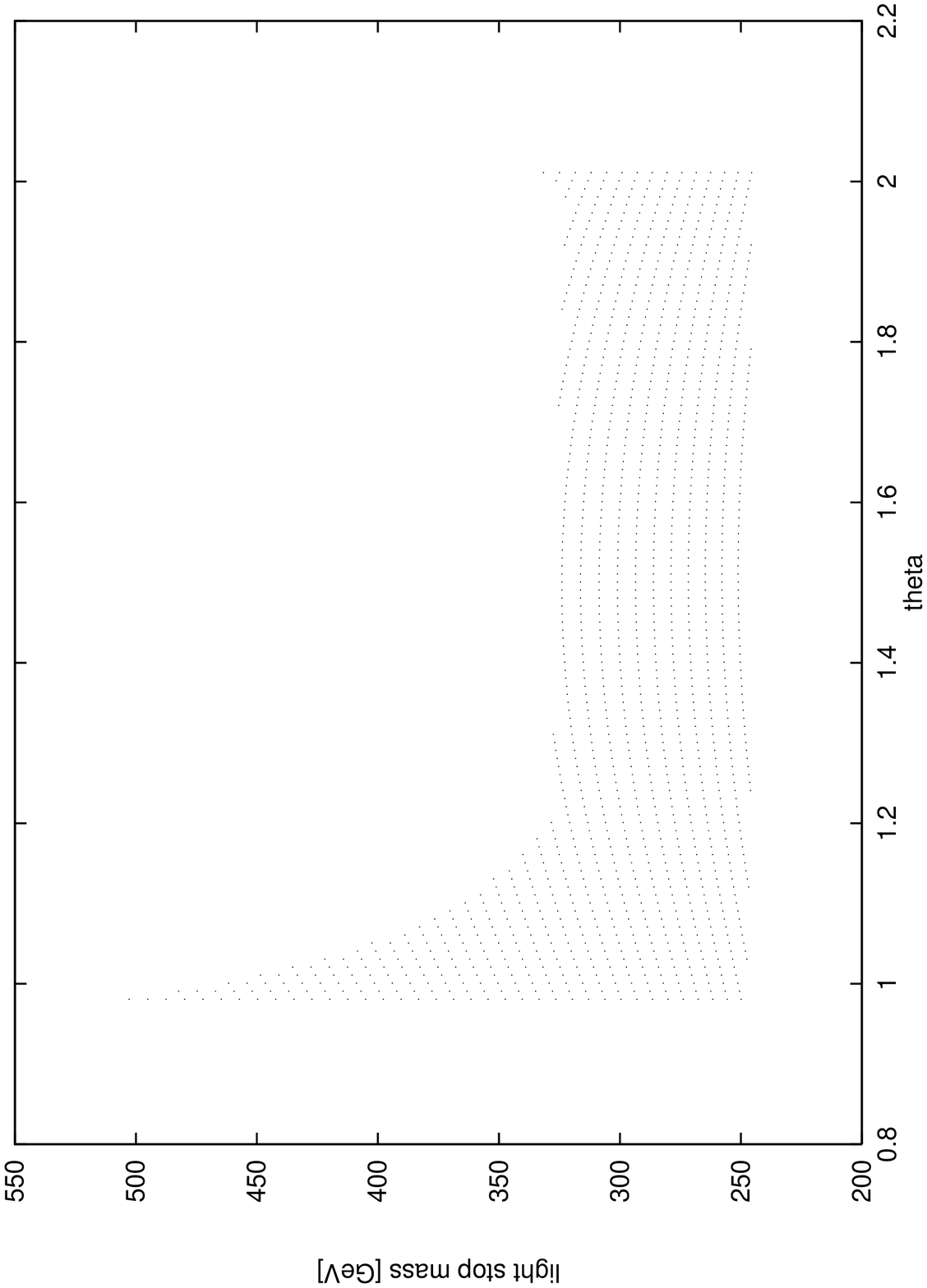}
\end{figure}
\begin{figure}
\epsfxsize=\hsize
\epsffile{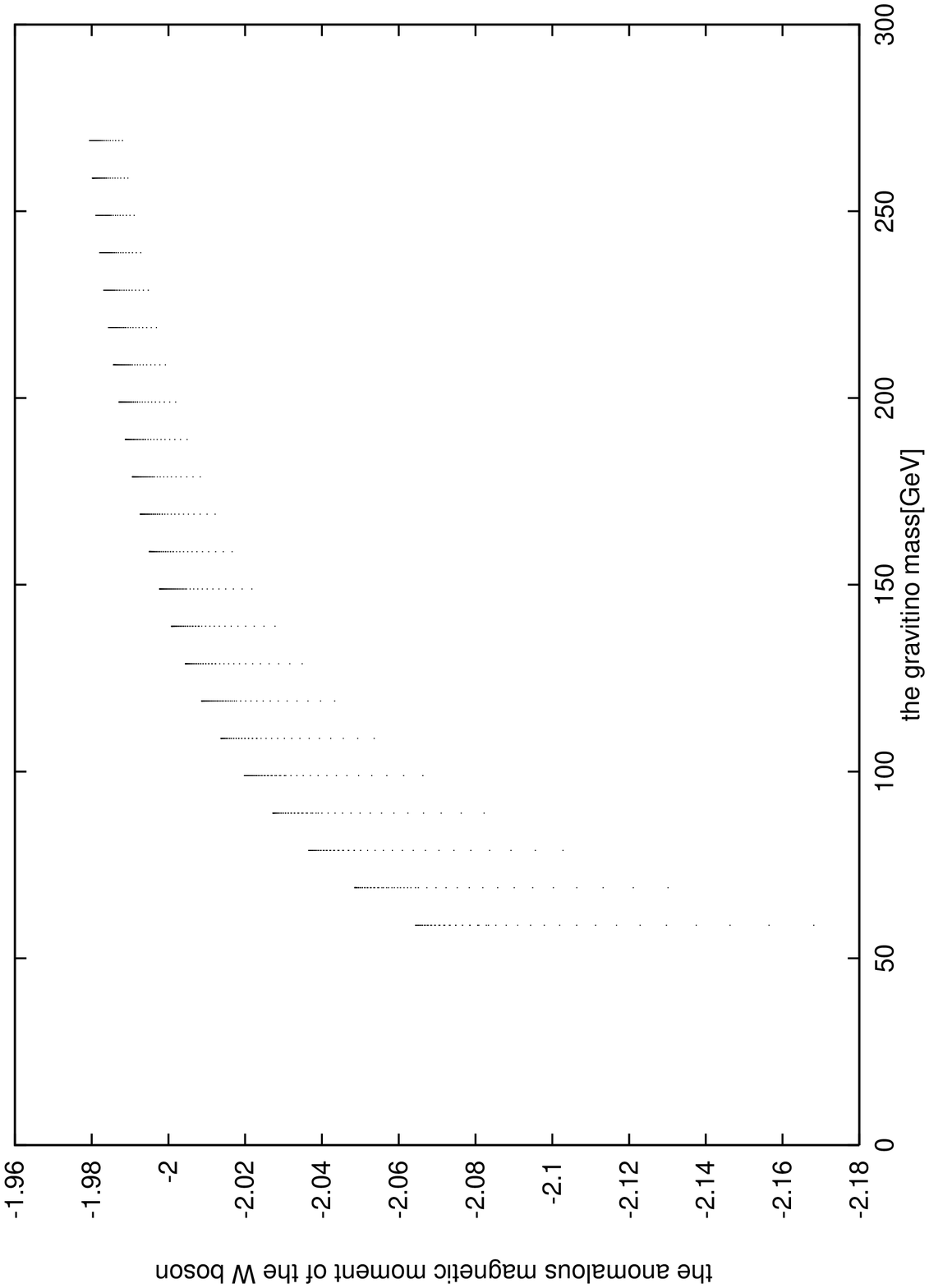}
\end{figure}

\end{document}